# A Fast Diagnosis Scheme for Distributed Small Embedded SRAMs

Baosheng Wang[1], Yuejian Wu[2] and André Ivanov[1]

[1]SOC Lab, Department of Electrical & Computer Engineering, University of British Columbia,
Vancouver, B.C., Canada, V6T 1Z4, {baosheng, ivanov}@ece.ubc.ca

[2]Nortel Networks, P.O. Box 3511, Station C, Ottawa, Ontario, Canada K1Y 4H7,
yuejian@nortelnetworks.com

**Abstract**

*This paper proposes a diagnosis scheme aimed at reducing diagnosis time of distributed small embedded SRAMs (e-SRAMs). This scheme improves the one proposed in [7, 8]. The improvements are mainly two-fold. On one hand, the diagnosis of time-consuming Data Retention Faults (DRFs), which is neglected by the diagnosis architecture in [7, 8], is now considered and performed via a DFT technique referred to as the "No Write Recovery Test Mode (NWRTM)". On the other hand, a pair comprising a Serial to Parallel Converter (SPC) and a Parallel to Serial Converter (PSC) is utilized to replace the bi-directional serial interface, to avoid the problems of serial fault masking and defect rate dependent diagnosis. Results from our evaluations show that the proposed diagnosis scheme achieves an increased diagnosis coverage and reduces diagnosis time compared to those obtained in [7, 8], with neglectable extra area cost.*

**Keywords**: *Distributed Small Embedded SRAMs, Memory Diagnosis, Data Retention Fault, SPC, PSC, Diagnosis Time*

## 1. Introduction

Currently, one of the System-on-a-Chip (SoC) paradigms is associated with a trend that an increasingly large number of small SRAMs are widely embedded for buffering data between different computational components [1]. As the embedded SRAMs (e-SRAMs) are increasingly dense, the ability for fault diagnose becomes more important. The diagnosis of such memories is not only important for locating the faulty cells such that repair can be done to improve the production yield, but also important for debugging the memory circuits for process improvement during the product development stage.

Difficulties with the diagnosis of distributed small e-SRAMs lie in the following: (i) External testers become increasingly incapable of diagnosing deeply embedded memories with limited external observability and controllability. Built-in-Self-Diagnosis (BISD) appears to be the only known cost-effective solution for such problems. (ii) For large numbers of relatively small e-SRAMs, the use of separate BISD controllers/circuitry can amount to unacceptable test area overheads. (iii) The spatial distribution occurring with several small e-SRAMs renders the routing of wires required for delivering patterns and analyzing the responses to be problematic, especially if a single BISD controller is to be shared by many SRAMs to keep overhead low. (iv) Their diagnosis time is dominated by the time for diagnosing Data Retention Faults (DRFs), which, in practice, are diagnosed by performing a read operation following a predetermined delay (e.g., 100 ms) [3]. In other words, the total diagnosis time for these small e-SRAMs is at least a couple of hundred milliseconds, regardless of the memory size/capacity and diagnosis methodology, e.g., parallel or sequential diagnosis.

To overcome the above challenges, previous work mainly focuses on developing diagnosis architectures that support parallel BISD with a single shared BISD controller [4-8]. The parallel diagnosis of distributed small e-SRAMs minimizes the diagnosis area overhead without negatively affecting the diagnosis coverage, while allowing a dramatic reduction in the total diagnosis time. However, the scheme in [4] only supports multiple small e-SRAMs of the same size, which is usually impractical in a real SoC. The architecture in [5, 6] has a separate general data background generator and control signal generator associated with each memory. This scheme is generally not feasible for diagnosing multiple distributed small e-SRAMs due to the routing and area penalty. The bi-directional serial interface used in [7, 8] for delivering patterns and collecting responses not only simplifies the routing from the BISD controller to the memories under diagnosis, but also solves the serial fault masking problem of the single-directional serial interface used in [9, 10]. Unfortunately, a March element with the bi-directional serial interface in [7, 8] can detect at most one fault. Thus, the memory diagnosis capability is dependent on the defect rate. This results in long diagnosis time, even under a reasonable defect rate. More importantly, all the previous work fails to consider the DRF which dominates the time for small e-SRAMs diagnosis. As a result, the diagnosis coverage is compromised and the diagnosis time reduction is overstated.

This paper proposes a new diagnosis scheme targeting total diagnosis time reduction for distributed small embedded SRAMs while still maintaining acceptable control signal routing complexity and corresponding area overhead. Our designs are based on those proposed in [7, 8]. We use a pair comprised of a Serial to Parallel Converter (SPC) and a Parallel to Serial Converter (PSC) to replace the bi-directional serial interface in [7, 8] for each e-SRAM. This avoids the problems of the serial fault masking and defect rate dependent diagnosis. We combine the proposed scheme with an effective design-for-test (DFT) technique known as "No Write Recovery Test Mode" (NWRTM) [11] for diagnosing DRFs without incurring any extra delay time. Together, these improvements yield a high coverage at the expense of a relatively short execution time and a low area overhead for diagnosing multiple, small distributed e-SRAMs.

The remainder of this paper is organized as follows. In Sec. 2, we briefly review the diagnosis architecture in [7, 8]. The detailed scheme designs for reducing the diagnosis time of distributed small SRAMs are proposed in Sec. 3. The diagnosis evaluations, i.e., diagnosis coverage analysis, diagnosis time comparison, and diagnosis area overhead estimation, are discussed in Sec. 4. Finally, Sec. 5 draws some conclusions.

## 2. Review of Previous Work

We begin by briefly reviewing the basics of the diagnosis architecture developed in [7, 8] for diagnosing distributed small e-SRAMs. To achieve low area overhead, a shared single BISD controller is adopted, which includes an address trigger to enable the address generators located local to each memory, a data background generator, and a control signal generator. The memory address generators are designed to be local to each memory to simplify routing. To overcome the data routing challenge and the serial fault masking problem arising from the single-directional serial interface [9, 10], a bi-directional serial interface is designed to deliver patterns and collect responses. These responses are routed back to the controller and compared with the expected values, bit by bit for each memory. Once a defective cell has been detected, it can be replaced with a spare cell if it is available. This diagnosis scheme is shown in Figure 1.

The DiagRSMarch algorithm used in [7, 8] is based on a March C- algorithm [12], but can detect all the faults covered by March CW [13] algorithm, which extends March C- by considering multiple data backgrounds.

The bi-directional serial interface in [7, 8] improves the single-directional serial scan circuit structures in [9, 10] during the data application and observation, such that no fault can be masked by a preceding fault and all the fault cells can be correctly identified. An example of this bi-directional interface technique is shown in Figure 2.

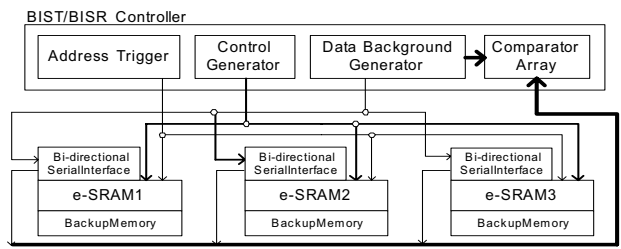

**Figure 1. The diagnosis scheme in [7, 8]**

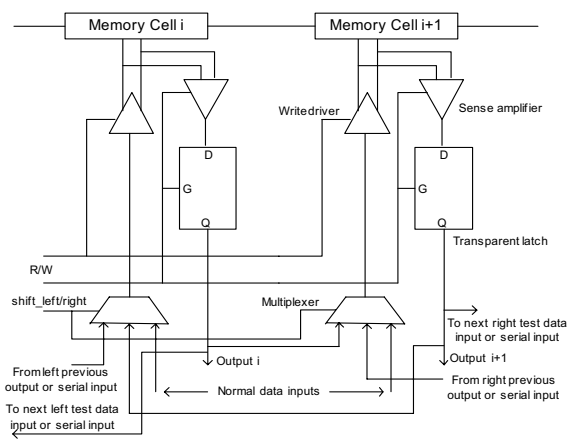

**Figure 2. Memory with bi-directional serial connections in the BISD mode**

With such a serial interface, at most one fault can be detected for each March element. In other words, the total diagnosis process is dependent on the defect rate. For a general manufacturing process with a reasonable defect rate, this diagnosis scheme will result in long diagnosis time. Moreover, the authors of [7, 8] do not consider the diagnosis of DRFs. Not considering DRFs limits diagnosis coverage and also mitigates the diagnosis time reduction achieved due to their proposed architecture since the time required for the diagnosis of DRFs usually dominates the small e-SRAM total diagnosis time.

## 3. The Fast Diagnosis Scheme

### 3.1 Architectural Overview

To improve the architecture in [7, 8] for shortened diagnosis time without imposing excessive area overhead, we propose a diagnosis scheme based on serial delivery but parallel application of patterns and serial analysis of responses. The patterns serially delivered to e-SRAMs are applied to these memories through the Serial to Parallel Converters (SPCs). The responses from each e-SRAM are routed back to the BISD controller in a serial fashion through the Parallel to Serial Converter (PSC). These responses are compared with the expected values, bit by bit for each e-SRAM by the comparator array. Once a defective cell has been detected, the diagnosis information, e.g., the faulty address, applied data background, etc., will be registered for on-chip repair or shifted out for off-line analysis. To avoid worsening the diagnosis signals routing problem, we design both the SPC and the PSC to reside locally to a memory under diagnosis. In other words, in our scheme, each e-SRAM possesses its own SPC and PSC. Like that in [7, 8], the memory address generator for each e-SRAM are also designed to be local to each memory to save the test address routing area. The global BISD controller is designed based on the largest (i.e., largest capacity) and the widest (largest IO number) e-SRAM (s).

A time-efficient method to diagnose DRFs is referred to as the "No Write Recovery Test Mode (NWRTM)" in [11]. We adopt this method here as well. Since this technique only requires a single control gate for the entire e-SRAMs to disable pre-charge bit lines during DRF diagnosis, an NWRTM signal is routed to all the memories. This signal is added into the control generator to enable the DRF diagnosis for all e-SRAMs.

For fair comparison, we use the March CW algorithm in [13] for the proposed diagnosis scheme. According to the above discussions, our proposed diagnosis scheme is shown in Figure 3.

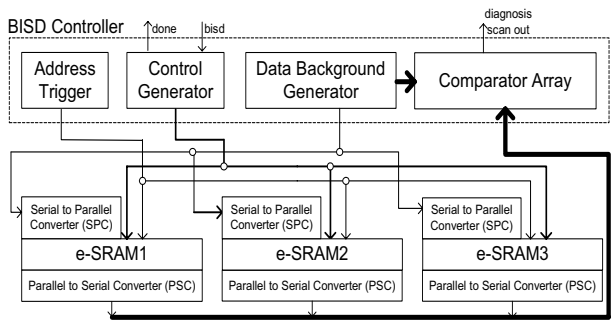

**Figure 3. The proposed diagnosis scheme**

The proposed scheme works as follows. Before each March element begins, a test pattern is serially delivered to all SPCs local to the e-SRAMs. Once the pattern delivery is complete, the controller triggers the local address generator to conduct a full March element before providing a new test pattern. During the read phase of each March element, once the memory responses are captured by the PSC, they are shifted back to the BISD controller, bit by bit, while the memory is in an idle or no-op mode. If a memory is not equipped with an idle or no-op mode, the memory is placed in a read mode however with data read ignored. Since our PSC shifting path does not involve the memories, there is no fault masking effect. The comparator array in the central controller compares these responses with the expected values bit by bit. Once a defective cell is found, the diagnosis information, e.g., failure addresses, data background, etc., will be either registered for on-chip repair or scanned out for off-line analysis.

Similarly to that described in [7, 8], a pattern is written to each address only once for the largest memory or memories. For smaller ones, however, the same pattern could be written on each address multiple times as the addresses wrap around. It should be pointed out that the responses obtained from a smaller e-SRAM will change as soon as the addresses wrap around for the first time, due to the read-modify-write operations of the March C-. Knowing when the addresses wrap around requires memory size information. This paper chooses to store this information in the BISD controller, just like that in [7, 8], so that the comparison in the BISD controller can tolerate those redundant read/write operations.

### 3.2 Serial to Parallel Converter (SPC)

In the proposed diagnosis scheme, SPCs receive the patterns delivered from the Data Background Generator of the BISD controller and apply these patterns to the corresponding e-SRAMs in parallel.

If the serial pattern from the Data Background Generator is shifted from the least significant bit (LSB) to the most significant bit (MSB) and the SPC also converts the patterns from the LSB to the MSB, different types of SPCs will be needed. E.g., if both pattern delivery and conversion are performed from LSB to MSB, the converted pattern for the widest e-SRAM and smaller one would be DP [c-1:0] and DP [c-1: c-c'], respectively, when the Data Background Generator delivers a pattern denoted as DP [c-1:0], where c and c' is the IO number of the widest e-SRAM and the narrower one respectively. This is because the first (c-c') bits of patterns for the narrower e-SRAMs are shifted out of the SPC and get lost. However, the expected patterns delivered to the narrower e-SRAMs should be DP [c'-1:0]. This mismatch may reduce diagnosis coverage.

To prevent the potential coverage loss, we design the pattern delivery and conversion for all memories according to the following: the serial pattern is shifted from the MSB to the LSB and its corresponding SPC coverts the patterns from the MSB to the LSB. As a result, the corresponding conversion from the Data Background Generator will also be modified as that from the MSB to LSB. With this appropriate design, all the patterns are correctly delivered in parallel to every small e-SRAM under diagnosis.

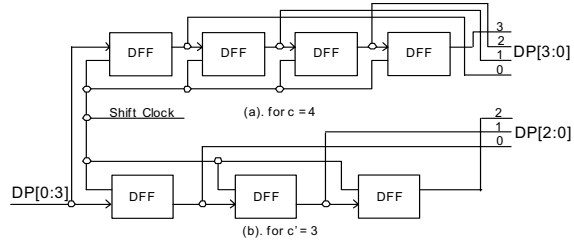

**Figure 4. Designs for pattern delivery and SPC**

A design example of two co-existing small e-SRAMs with c = 4 and c'= 3 is shown in Figure 4 (a) and (b), respectively. Obviously, the widest e-SRAM has an IO number of 4 while the narrower one has an IO number of 3 in this example. It should be pointed out that a new diagnosis pattern is delivered to the SPC only once just before each March element begins.

### 3.3 Parallel to Serial Converter (PSC)

The proposed PSCs collect diagnosis responses from each e-SRAM in parallel and convert them into serial ones. These serialized response sequences are shifted back to the BISD controller from the LSB to MSB. Although the response analysis for each e-SRAM is in a serial manner, the response sequences from all the memories are analyzed in parallel.

Since all the PSCs are independent of each other, they can be designed to be the same for each e-SRAMs, i.e., to go from LSB to MSB. To separate the memory output from the shifting components, a scan type of DFFs are adopted. An example PSC is shown in Figure 5.

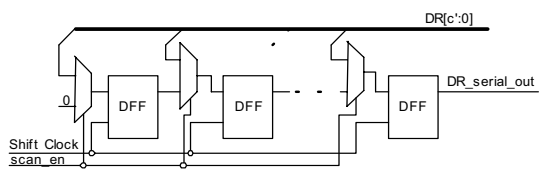

**Figure 5. Design for a general PSC**

In Figure 5, the diagnosis responses are first captured into c' registers in parallel. This is followed by the memory entering an idle mode when shifting the memory responses back to the BISD controller for evaluation. If a memory does not have an idle mode, we can place the memory in read mode with read data ignored during the shift operation of the PSC. Since the memory does not interfere with shift operation of the PSC, there will be no negative impact on diagnosis coverage due to the extra read operations. However, an extra scan_en signal is required to control the capture of memory test response and the shift or serialization of the captured test response. It should be pointed out that the serialization operation of the PSC with the memory in an idle mode does not compromise at-speed diagnosis coverage. This is because in the read-modify-write operations used in the March C-, e.g., (R0 W1), the only signals that change after the R0 and before the W1 are the read/write enable (WEN) and data inputs. As long as we ensure that the WEN and data inputs do not change until the last shift operation in the PSC, the shift operation does not change at-speed coverage of the WEN decoding and data input circuitry.

### 3.4 Diagnosis of Data Retention Faults

To diagnose DRFs, we adopt a previously proposed low-penalty DFT technique from [11], referred to as "No Write Recovery Test Mode (NWRTM)".

Like the methodology in [14] and [15], in NWRTM, a special write cycle is created to distinguish a good cell from a faulty cell when subjected to a DRF caused by an open defect on the pull-up PMOS. A typical 6T SRAM cell with storage node A and complementary storage node B, shown in Figure 6, is used to illustrate the differences between the specifically designed "No Write Recovery Cycle (NWRC)" and a normal write cycle. The bitline precharge circuit of NWRTM is also shown in Figure 6, where the signal NWRTM is used to switch between the NWRC and a normal write cycle. In Figure 6, during a normal W1 cycle, node B is pulled down by the bitline BLb that is driven to "true" GND by the write control logic; and node A is pulled up by its pull-up PMOS. Here, "true" GND means that the node is driven to the GND voltage level by an active device. Due to the latch mechanism of the memory cell, the cell flips its value from "ZERO" to "ONE" as long as the voltage difference between nodes A and B reaches a threshold.

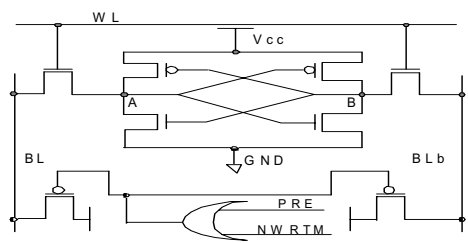

**Figure 6. A typical 6T SRAM cell and its pre-charge control circuits for NWRTM**

In [14] and [15], by setting the bitlines BL and BLb to a given voltage level between Vcc and GND during the write operation, a good cell fails to flip while a faulty cell does. Similarly, we set the voltage level of BL and BLb to "float" GND and "true" GND respectively. Here, "float" GND means the node voltage level is at GND but not actively driven by any device. This causes the opposite result, i.e., a good cell succeeds at flipping its logic value while a faulty cell fails to do so. For the above example, a good cell has no problem writing a "ONE" because node B can be pulled down by the bitline BLb and the cell can flip to "ONE" due to the latch mechanism. However, a faulty cell subject to a DRF fails to flip because the voltage level of node A never exceeds that of node B. The voltage level of node A always remains at GND since (i) lacking the PMOS or path to the supply rail, node A cannot be pulled to "ONE" regardless of how low the node B voltage level reaches. Thus, the latch in this faulty cell malfunctions; and (ii) there are no charge sharing effects with bitline BL because it is set at "float" GND. GND is the lowest achievable voltage level and node A remains at GND. Consequently, the voltage level of node A never exceeds that of node B and the faulty cell fails to flip. Therefore, DRFs are detected under NWRTM.

Like a normal write operation, a NWRTM write operation can successfully write a good cell and may fail to write a defective cell causing DRFs. Therefore, the NWRTM can be merged with any March test by simply adding two extra NWRCs just before the normal write [11]. Other DFT techniques do not share this advantage. Hence, NWRTM is the best in terms of test time for DRFs among all existing DFT techniques.

## 4. Evaluations

### 4.1 Diagnosis Coverage Analysis

Compared with the diagnostic scheme in [7, 8], the proposed scheme simply replaces the bi-directional serial interface with a pair of SPC and PSC. All the other components in [7, 8] are preserved. In terms of diagnostic coverage, the adoption of the March CW in this paper provides the same coverage as the serialized March C- used in [7, 8]. However, due to the use of the NWRTM, the proposed scheme achieves additional coverage of DRFs. As a result, the diagnosis coverage of the proposed diagnosis scheme is increased compared with that of those in [7, 8] because its diagnosis capacities in DRFs and other defects not causing faulty logical behaviors but possibly causing reliability problems.

### 4.2 Diagnosis Time Comparison

Since the DRFs diagnosis is not considered in [7, 8], the reported diagnosis time reduction can be considered to be optimistic. With our proposed diagnosis scheme, the diagnosis time is much less than that in [7, 8].

The DiagRSMarch algorithm in [7, 8] is based on right-shift operational RSMarch with extra March elements that include both left-shift operations and checkboard patterns. Therefore, assuming the largest/widest e-SRAM under diagnosis has a capacity of $n$ words and IO number of $c$, the diagnosis time of the DiagRSMarch algorithm in [7, 8] without considering DRFs diagnosis is

$$T_{[7,8]} = 17knct + 9nct = (17k+9)nct \qquad (1)$$

where $t$ is the diagnosis clock period (ns) and $k$ is iteration number of M1 elements required in [7,8].

Without DRFs diagnosis, the diagnosis time of the selected March CW algorithm for our proposed diagnosis scheme can be calculated as

$$T_{proposed} = \{(5n+5c+5n(c+1))+(3n+3c+2n(c+1))\lceil \log_2 c \rceil\}t \qquad (2)$$

where $(5n+5c+5n(c+1))$ is the complexity of a parallel March C- algorithm with our diagnosis scheme; $(3n+3c+2n(c+1))\lceil \log_2 c \rceil$ represents the complexity due to the added March element in March CW for detecting intra-word and column decoder faults.

The diagnosis time reduction we achieve with the proposed scheme is given by the following

$$R = \frac{T_{[7,8]}}{T_{proposed}} = \frac{(17k+9)nc}{(10n+5c+5nc)+(5n+3c+2nc)\lceil \log_2 c \rceil} \qquad (3)$$

Although not obvious, the reduction factor $R$ will always exceed one in practice because the iteration number $k$ is always much larger than one.

If the DRFs are considered, the extra diagnosis time for DRFs when using the diagnosis architecture in [7, 8] includes $8k$ units of extra complexities (i.e., $(w0/r0)_{R+L}$, $(w1/r1)_{R+L}$) and 200ms delay time. In comparison, the proposed scheme requires only 2 units of extra test complexities (i.e., Nw0/Nw1 in [11]) for DRFs diagnosis. In this case, the diagnosis time ratio $R$ can be calculated as shown in Equation (4):

$$R = \frac{T_{[7,8]} + 8knct + 2 \times 10^8}{T_{proposed} + (2n+2c)t} \qquad (4)$$

where $t$, $T_{[7,8]}$ and $T_{proposed}$ are in ns.

From equation (4), the reduction ratio due to the proposed diagnosis scheme could be extremely high when DRFs diagnosis is included.

To quantitatively investigate the diagnosis time reduction, we use a case study in [16] as the benchmark e-SRAMs,

where *n = 512, c = 100 and t = 10ns*. We assume that 1% of the memory cells are defective and all four different defect types in [8] occur with equal likelihood. From [8], the maximum numbers of the total faults for each of the benchmark e-SRAMs in [16] would be 256. Since the M1 element in [7, 8] can cover 75% of those faults and each iteration of the M1 element can identify at most two faults, the minimum iteration number *k* can be calculated to be (256*0.75/2) = 96. Using these assumptions, we found this diagnosis time reduction factor *R*, without considering DRFs, is at least 84. If DRFs are considered, *R* for the e-SRAMs in [16] can be at least 145.

### 4.3 Area Overhead Estimations

Area overhead of the proposed scheme is evaluated according to the required number of transistors to implement the scheme and the number of global wires.

Compared with the designs in [7, 8], the proposed scheme adds only one extra global wire for the control of the PSC.

In terms of transistor count, the bi-directional serial interface [7, 8] actually includes a set of 4:1 multiplxers and latches. In the proposed scheme, a SPC and a PSC together require two sets of shift registers and 2:1 multiplexers, one for selecting between normal and test inputs and the other for the scan DFFs in PSC. In terms of transistor count, we find that a D-flip-flop is equivalent to two 6T SRAM cells while a latch is equivalent to one 6T SRAM cells. Therefore, this total area overhead extra to [7, 8] is three 6T SRAM cells per bit. Fortunately, this extra area can be neglectable in practice. For example, this area overhead is around 1.8% for the benchmark e-SRAMs in [16] when applying both that in [7, 8] and the proposed diagnosis scheme.

### 5. Conclusions

The major challenge of diagnosing distributed small e-SRAMs is not only the diagnosis area overhead in terms of diagnosis circuits and wires routing, but also the diagnosis time under high coverage requirements. This paper presented a significant improvement on the diagnosis architecture described in [7, 8]. By replacing the bi-directional serial interface used in [7, 8], the proposed scheme greatly reduced diagnosis time. By adopting a previous DFT technique known as "NWRTM" to detect data retention faults, the proposed scheme achieved better coverage. Compared with those in [7, 8], the evaluation results indicate that the diagnosis time under a reasonable 1% defect rate environment is reduced by a factor of at least 84 with 1.8% of the total cells area extra to that in [7, 8].

### 6. References

[1] A. Bommireddy, et. al., "Test and debug of networking SoCs – a case study", *proceedings of IEEE 18th VLSI test Symposium (VTS00)*, pp. 121- 126, Montreal, Apr. 2000.

[2] C. Selva, et. al., "A programmable built-in self-diagnosis for embedded SRAM", *Proceedings of the 2004 International Workshop on Memory Technology, Design and Testing (MTDT04)*, pp. 84 - 89, 2004

[3] B. Wang, et. al., "Reducing embedded SRAM test time under redundancy constraints", *Proceedings of the 22nd IEEE VLSI Test Symposium (VTS04)*, pp. 237 - 242, 2004.

[4] L. M. Deng, et. al., "A parallel built-in self-diagnosis scheme for embedded memory", *the 2004 IEEE International Workshop on Memory Technology, Design, and Testing (MTDT04)*, pp. 65 - 69, Aug. 9-10, 2004.

[5] C. W. Wang, et. al., "A built-in self-test and self-diagnosis scheme for heterogeneous SRAM clusters", *Proceedings of 10th Asian Test Symposium*, pp. 103 - 108, 2001.

[6] R. C. Aitken, "A modular wrapper enabling high speed BIST and repair for small wide memories", *the International Test Conference (ITC04)*, pp. 997 – 1005, Oct. 26-28, 2004.

[7] D. C. Huang, et. al., "A parallel built-in self-diagnostic method for embedded memory buffers", *the International Conference on VLSI Design*, pp. 397-402, 2001.

[8] D. C. Huang and W. B. Jone, "A parallel built-in self-diagnostic method for embedded memory arrays", *IEEE Transactions on Computer-Aided Design of Integrated Circuits and Systems*, Vol. 21, Issue 4, pp. 449-465, 2002.

[9] B. Nadeau-Dostie, et. al., "A serial interfacing technique for built-in and external testing of embedded memories", *Proceedings of the IEEE 1989 Custom Integrated Circuits Conference*, pp. 22.2/1- 22.2/5, May 1989.

[10] B. Nadeau-Dostie, et. al., "Serial interfacing for embedded-memory testing", *IEEE Design & Test of Computers*, Vol. 7, No. 2, pp. 52 -63, April 1990.

[11] J. Yang, et. al., "Open defects detection within 6T SRAM cells using a No Write Recovery Test Mode", *the Proceedings of International Conference on VLSI Design 2004*, pp. 493 – 498, Jan. 5-9, 2004.

[12] M. Marinescu, "Simple and efficient algorithms for functional RAM testing", *IEEE Proceedings of International Test Conference*, pp. 236-239, 1982.

[13] C. F. Wu, et. al., "RAMSES: a fast memory fault simulator", *International Symposium on Defect and Fault Tolerance in VLSI Systems*, pp. 165 – 173, 1999.

[14] A. Meixner, and J. Banik, "Weak write test mode: an SRAM cell stability design for test technique", *Proceedings of International Test Conference*, pp. 309 – 318, 20-25 October 1996.

[15] V. H. Champac, et. al., "Bit line sensing strategy for testing for data retention faults in CMOS SRAMs", *Electronics Letters*, Vol. 36, Issue 14, pp. 1182-1183, 6 July 2000.

[16] B. Wang, et. al., "Designs for Reducing Test Time of Distributed Small Embedded SRAMs", in *IEEE International Symposium on Defect and Fault Tolerance in VLSI Systems*, pp. 120 - 128, Cannes, France, Oct. 10-13, 2004.